\documentclass[twocolumn]{aastex63}

\shorttitle{Accelerating the Venusian Atmospheric Evolution}
\shortauthors{Stephen R. Kane et al.}

\begin{document}

\title{Could the Migration of Jupiter have Accelerated the Atmospheric
  Evolution of Venus?}

\author{Stephen R. Kane}
\affiliation{Department of Earth and Planetary Sciences, University of
  California, Riverside, CA 92521, USA}
\email{skane@ucr.edu}

\author{Pam Vervoort}
\affiliation{Department of Earth and Planetary Sciences, University of
  California, Riverside, CA 92521, USA}

\author{Jonathan Horner}
\affiliation{Centre for Astrophysics, University of Southern
  Queensland, Toowoomba, QLD 4350, Australia}

\author{Francisco J. Pozuelos}
\affiliation{Space Sciences, Technologies and Astrophysics Research
  (STAR) Institute, Universit\'e de Li\`ege, All\'ee du 6 Ao\^ut 19C,
  B-4000 Li\`ege, Belgium}
\affiliation{Astrobiology Research Unit, Universit\'e de Li\`ege,
  All\'ee du 6 Ao\^ut 19C, B-4000 Li\`ege, Belgium}


\begin{abstract}

In the study of planetary habitability and terrestrial atmospheric
evolution, the divergence of surface conditions for Venus and Earth
remains an area of active research. Among the intrinsic and external
influences on the Venusian climate history are orbital changes due to
giant planet migration that have both variable incident flux and tidal
heating consequences. Here, we present the results of a study that
explores the effect of Jupiter's location on the orbital parameters of
Venus and subsequent potential water loss scenarios. Our dynamical
simulations show that various scenarios of Jovian migration could have
resulted in orbital eccentricities for Venus as high as 0.31. We
quantify the implications of the increased eccentricity, including
tidal energy, surface energy flux, and the variable insolation flux
expected from the faint young Sun. The tidal circularization timescale
calculations demonstrate that a relatively high tidal dissipation
factor is required to reduce the eccentricity of Venus to the present
value, which implies a high initial water inventory. We further
estimate the consequences of high orbital eccentricity on water loss,
and estimate that the water loss rate may have increased by at least
$\sim$5\% compared with the circular orbit case as a result of orbital
forcing. We argue that these eccentricity variations for the young
Venus may have accelerated the atmospheric evolution of Venus toward
the inevitable collapse of the atmosphere into a runaway greenhouse
state. The presence of giant planets in exoplanetary systems may
likewise increase the expected rate of Venus analogs in those systems.

\end{abstract}



\section{Introduction}
\label{intro}

The current state of the Venusian atmosphere and the pathway through
which it arrived there is an exceptionally complicated topic. Numerous
studies have provided insights into the climate evolution of Venus and
discussed primary influences on the atmospheric dynamics
\citep{bullock1996,taylor2009,taylor2018}, including obliquity
variations \citep{barnes2016a}. The evolutionary history of the
atmosphere of Venus, and its potential divergence from a temperate
``Earth-like'' climate, depends heavily upon assumptions regarding the
initial conditions. For example, \citet{hamano2013} proposed that
Venus may have never had surface liquid water oceans due to an
extended magma surface phase. \citet{ramirez2020b} suggested that a
high nitrogen inventory on a young planet could lead to a delay in
runaway greenhouse scenarios, and that such a scenario might have
caused Venus to transition directly into a runaway state without the
need for a moist greenhouse. Alternatively, some models suggest that
Venus may have had temperate surface conditions that allowed the
persistence of surface liquid water until as recently as $\sim$0.7~Ga
\citep{way2016}, depending upon assumptions regarding rotation rates
and convection schemes
\citep[e.g.,][]{leconte2013c,ramirez2018d}. Such potential for past
Venusian surface habitability has been the basis for defining the
empirically derived inner edge of the ``Habitable Zone''
\citep{kasting1993a,kopparapu2013a,kopparapu2014,kane2016c}. The
connection to planetary habitability has further fueled the relevance
of Venus to refining models of exoplanets \citep{kane2019d}, both in
terms of studying atmospheric chemistry
\citep{schaefer2011,ehrenreich2012a} and detection prospects for
potential Venus analogs \citep{kane2013d,kane2014e,ostberg2019}.

In the consideration of climate evolution, the orbital parameters of a
planet can play a key role in the energy budget distribution over the
surface of the planet \citep{kane2017d}. In particular, it has been
demonstrated that the orbital eccentricity can have significant
consequences for the climate evolution of terrestrial planets
\citep{williams2002,dressing2010,kane2012e,bolmont2016a,way2017a,palubski2020a}. Overall
planetary system architectures can also play a role, such as the
effect of Jupiter on impact rates
\citep[e.g.,][]{horner2008a,horner2009a,horner2012b} and refractory
elemental abundance \citep[e.g.,][]{horner2009b,desch2018} in the early
inner solar system. \citet{correia2012a} showed that the eccentricity
of planetary orbits can be increased by the excitation effects of
outer planets that exceed the dampening effects of tidal heating. For
those planets where the eccentricity contributes to significant tidal
heating, the additional surface energy flux can trigger a runaway
greenhouse for an otherwise temperate terrestrial planet
\citep{barnes2013a}. Furthermore, the current rotation rate of Venus
appears to be impacted by eccentricity and resulting solar tidal
torques \citep{ingersoll1978b,bills2005e,green2019}, in addition to
interactions between the atmosphere and topography
\citep{fukuhara2017a,navarro2018}.

At the present epoch, Venus has the lowest orbital eccentricity ($e_V
= 0.006$) of the solar system planets and is also in a post-runaway
greenhouse state. However, both of these aspects of Venus have a time
evolution component. For example, the eccentricity of the Venusian
orbit has both increased due to perturbations from the other planets
and decreased due to tidal dissipation. In this paper, we explore the
effects of possible giant planet migration on the Venusian orbital
eccentricity in the early history of the solar system and the
possible impacts on the climate evolution of the planet. In
Section~\ref{dynamics} we provide the details of our dynamical
analysis and the range of eccentricities that Venus could have
attained, depending on the location of Jupiter. We further calculate
tidal dissipation timescales and show that a significant initial
volatile inventory can greatly decrease circularization
time frames. Section~\ref{con} discusses the consequences of a high
Venusian eccentricity, including tidal effects, insolation, and water
loss rates. Section~\ref{climate} describes the relative effect of
eccentricity evolution for both Earth and Venus, including water loss
scenarios. In Section~\ref{implications} we discuss the potential
implications of this analysis for terrestrial planets in systems with
giant planets, and we provide concluding remarks in
Section~\ref{conclusions}.


\section{Evolution of the Venusian Orbit}
\label{dynamics}

\subsection{Migration and Formation Scenarios}
\label{migration}

As we have learned more about the solar system, it has become clear
that the giant planets must have migrated during their formation and
evolution, before settling at their current locations. Evidence for
that migration abounds in the system's various small body
populations. These include the sculpting of the asteroid belt and
orbital distribution of the Jovian Trojans \citep[e.g.,][and
  references therein]{morbidelli2005,minton2009}, and the excitation
of the Plutinos and distribution of objects beyond Neptune
\citep[e.g.,][]{malhotra1993b,malhotra1995b,hahn2005b,levison2008a,lykawka2009b,lykawka2010a}. Although
significant giant planet migration is now firmly established, there
remains debate over both the extent of that migration, its smoothness
or chaoticity, and its timing. Migration models have ranged from the
late ($\sim$700~Myr after planet formation) chaotic interactions of
the Nice Model \citep[e.g.,][]{gomes2005b,morbidelli2005} to the early
but dramatic ''Grand Tack'' migration of Jupiter and Saturn, in which
it is suggested that Jupiter might have migrated inward to approach
the current orbit of Mars, before tacking outward to reach its current
location \citep[e.g.,][]{walsh2011c,raymond2014a,nesvorny2018c}, a
process that would have had a significant impact on the hydration of
the inner solar system \citep[e.g.,][]{obrien2014a,raymond2017b}. In
particular, the predominant version of the Grand Tack model assumes a
disk-dominated migration of Jupiter that was completed before the
terrestrial planets formed
\citep[e.g.,][]{chambers2014c}. Alternatively, models of
planetesimal-driven migration of the giant planets can occur over much
longer time scales, far beyond the dissipation of the disk required by
the Grand Tack model \citep{hahn1999,malhotra2019}. Further migration
models consider the possibility of non-uniform, stochastic migration
\citep[e.g.,][]{morbidelli2010a,nesvorny2018d}, whereas other models
consider more sedate, smoother migration
\citep[e.g.,][]{lykawka2009b,lykawka2010a,pirani2019a}. Additional
complications include evidence via isotopic measurements from
meteorites that suggest terrestrial protoplanets completed accretion
quickly and while the gas was still present \citep{schiller2018}, and
the prevalence of proto-atmospheres for terrestrial planets whilst
still present in the gas phase \citep{mai2020}. A summary of migration
and formation models is provided by \citet{raymond2020}.

Regardless of the true nature of the giant planet migration, the past
few decades have seen extensive investigation of the dynamical
evolution of the solar system
\citep[e.g.,][]{laskar1988b,duncan1993,levison2003a,batygin2008,brasser2009,zeebe2015a}. In
addition to investigations of an early solar system, dynamical
simulations by \citet{laskar1988b} considered the current stability of
the solar system, finding that the orbital eccentricities of both
Venus and Earth remain largely circular over several million year
timescales, based on their present orbital parameters. As described
previously, several theories regarding the formation and migration of
the giant planets, Jupiter and Saturn, suggest that there may have
been periods during which they were significantly closer to the Sun,
including the Nice model \citep{gomes2005b,tsiganis2005b} and the
complementary Grand Tack model \citep{walsh2011c,walsh2012c}. Given
the aforementioned variety in migration theories discussed in the
literature, the effects of that migration on the inner solar system
are now the subject of investigation. It has been suggested that the
dynamical stability of the terrestrial planets may serve as a useful
discriminant between the different Jupiter migration models
\citep{nesvorny2018c}, with such models considered likely to help
explain the relatively high eccentricity of Mercury's orbit
\citep{roig2016}.

The purpose of this work is not to evaluate the relative merits of
these and other similar models, but rather to assess the impact of
Jupiter's location on the orbital dynamics of the inner
planets. However, it is worth noting that our model largely depends on
Jupiter's migration occurring after Venus's formation is complete or a
proto-Venus that retains eccentricity from formation
processes. Numerous models described herein allow for these
possibilities, and so the subsequent discussion describes the results
of a detailed dynamical simulation and the implications for the
orbital eccentricity of Venus.


\subsection{Dynamical Simulation}
\label{sim}

\begin{figure}
  \includegraphics[angle=270,width=8.5cm]{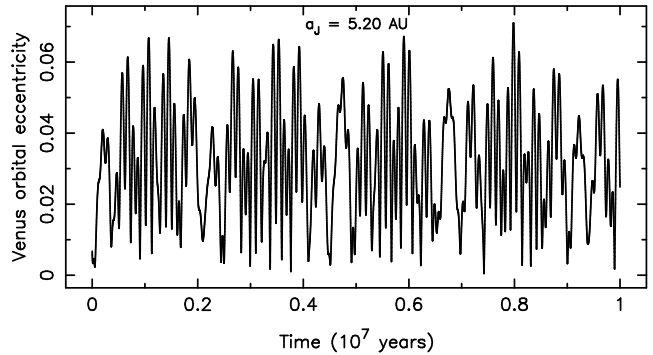}
  \caption{Orbital eccentricity of Venus as a function of time for the
    present solar system configuration ($a_J = 5.20$~AU). Time zero
    corresponds to the current epoch when the eccentricity resides at
    a periodic low point. The eccentricity peaks at a value of 0.07
    and exhibits high-frequency and low-frequency variations, with
    periods of $4.1 \times 10^5$~years and $2.2 \times 10^6$~years
    respectively.}
  \label{eccfig1}
\end{figure}

The dynamical simulations required for this study were conducted as
part of a broader investigation of the degree to which Jupiter's
semi-major axis and eccentricity drive the time-variability of
dynamical behavior in the solar system.  A detailed description of
the dynamical simulations can be found in \citet{horner2020a}, which
concentrated on the effect of Jupiter on the Milankovitch cycles of
the Earth. Here we provide a brief description of the simulations and
the components of the simulation that are utilized in this study.

The full suite of simulations was constructed using $N$-body
integrations calculated using the Hybrid integrator within the {\sc
  Mercury} $N$-body dynamics package \citep{chambers1999}. The
simulations included the eight major planets of the solar system and
incorporated the current orbital elements extracted from the Horizons
DE431 ephemerides \citep{folkner2014}. The full suite of simulations
covered a large grid of Jupiter semi-major axes (3.2--7.2~AU) and
eccentricities (0.0--0.4), totaling 159,201 simulations. Each
simulation was run for $10^7$ simulation years, with a time step of 1
day to ensure perturbation accuracy, and first-order post-Newtonian
relativistic corrections were accounted for \citep{gilmore2008}. The
output of each simulation was recorded with a time step of 1,000
years. Simulations were halted early if any of the planets became
``lost'', including collisions with the Sun or each other, or moving
beyond a heliocentric distance of 40~AU.

\begin{figure*}
  \begin{center}
    \includegraphics[angle=270,width=16.0cm]{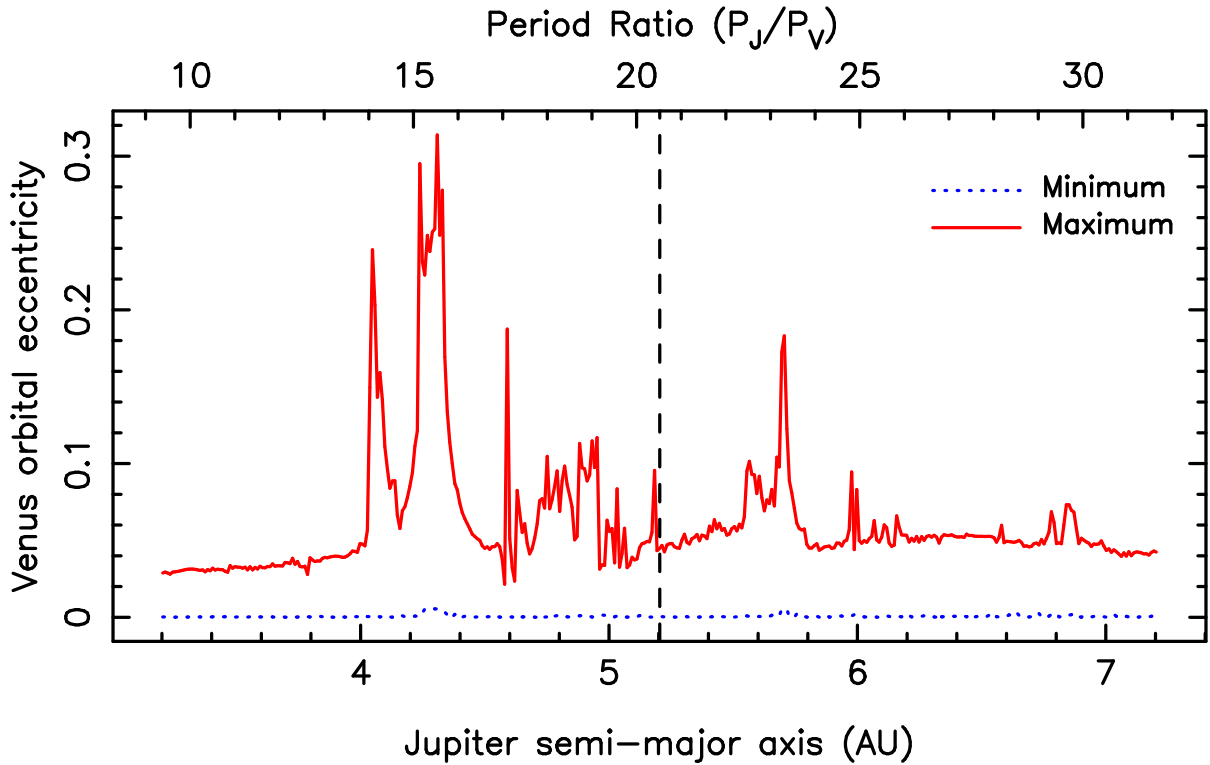}
  \end{center}
  \caption{Orbital eccentricity of Venus as a function of the
    semi-major axis of Jupiter. The dotted (blue) and solid (red)
    lines indicate the minimum and maximum eccentricities respectively
    that occur during the simulation for the shown Jupiter semi-major
    axis. The vertical dashed line indicates the present location of
    Jupiter. The top axis shows the corresponding period ratio between
    Jupiter ($P_J$) and Venus ($P_V$).}
  \label{eccfig2}
\end{figure*}

Our study of Venus uses 399 of the total simulation set, and includes
starting orbital parameters of Jupiter with zero eccentricity in the
semi-major axis range $a_J = 3.2$--7.2~AU. We restricted this study to
those simulations for which the initial eccentricity of Jupiter's
orbit was zero, in order to minimize assumptions regarding the
potential migration scenarios for Jupiter and to focus our work on the
the influence of the giant planet's location alone.


\subsection{Eccentricity of the Venusian Orbit}
\label{orbit}

As noted in Section~\ref{intro}, the present orbital eccentricity of
Venus is the lowest of all solar system major planets, with a value of
0.006. The results from our simulation for the case of a Jupiter
semi-major axis of 5.20~AU and eccentricity of 0.048 (present
configuration) are shown in Figure~\ref{eccfig1}. These results
demonstrate that the present orbital eccentricity of Venus is at a
periodic low point, with the maximum value of the periodic
eccentricity oscillations of $\sim$0.07 in agreement with the results
from \citet{laskar1988b} and \citet{bills2005e}. A Fourier analysis of
the data represented in Figure~\ref{eccfig1} revealed that the
eccentricity variations exhibit high-frequency periodicity of $4.1
\times 10^5$~years and a low-frequency periodicity of $2.2 \times
10^6$~years, primarily a result of perturbations from the Earth and
Jupiter. For comparison, the eccentricity of the Earth varies in the
range 0.0034--0.058 and is currently 0.0167 \citep{laskar1988b}, the
variations of which were also validated by our model
\citep{horner2020a}.

The relatively mild variations in the orbital eccentricity of Venus
exhibited in Figure~\ref{eccfig1} are sensitive to the location of
Jupiter. Using the results of our extensive suite of dynamical
simulations, we extracted the minimum and maximum orbital
eccentricities attained by Venus for the full range of Jupiter
semi-major axis values. These results are shown in
Figure~\ref{eccfig2}, where the dotted (blue) and solid (red) lines
indicate the minimum and maximum eccentricity, respectively, for the
given Jupiter semi-major axis. The present location of Jupiter is
shown as a vertical dashed line. The top axis provides the equivalent
period ratio between Jupiter ($P_J$) and Venus ($P_V$) as a guide for
possible resonance sources of instability.

The eccentricity data shown in Figure~\ref{eccfig2} demonstrate the
non-trivial relationship between the location of Jupiter and the
orbital variability of Venus. It also shows that there are numerous
locations at which the influence of Jupiter would cause the range of
eccentricity values for Venus to be significantly larger than the
present range shown in Figure~\ref{eccfig1}. For example, at $a_J =
5.18$~AU, only 0.02~AU away from its present location, the maximum
eccentricity of Venus rises to 0.10. There are also Jupiter locations
farther from its present position that result in large increases in
the maximum eccentricity, most particularly at $a_J = 5.71$~AU, where
the maximum Venusian eccentricity rises to 0.18.

However, it is evident from the eccentricity data that the most
powerful perturbations to the Venusian orbit occur when Jupiter is
located in the vicinity of $a_J \sim 4.3$~AU. The maximum Venusian
eccentricity of 0.31 occurs at a Jupiter semi-major axis of $a_J =
4.31$~AU. It should be noted that not all of the simulations
represented in Figure~\ref{eccfig2} are dynamically viable for the
full $10^7$ year duration. Moreover, as noted by \citet{horner2020a},
placing Jupiter at $\sim$4.3~AU produced instability within the system
at all Jovian eccentricities, even zero. Even so, those models of
planetary migration that invoke extreme excursions in Jupiter's
orbital elements (such as the Grand Tack model) require the planet to
have moved through this location, albeit probably on timescales that
allow unstable regions to be effectively circumnavigated, and so such
locations and resulting eccentricities are potentially valid regions
of exploration.

\begin{figure}
  \includegraphics[angle=270,width=8.5cm]{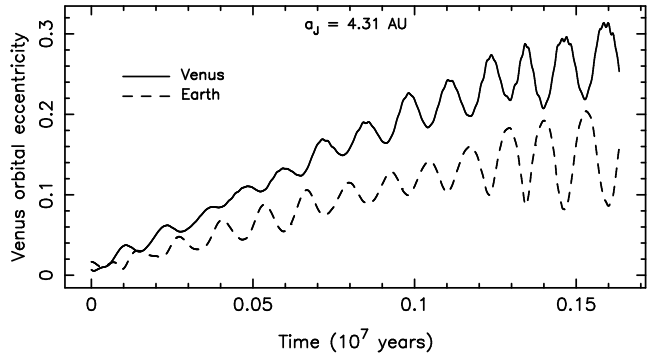}
  \caption{Orbital eccentricity of both Venus (solid line) and Earth
    (dashed line) as a function of time when Jupiter is located at a
    semi-major axis of 4.31~AU. This particular simulation halted
    early due to the system instability criteria being met at $1.6
    \times 10^6$~years (see Section~\ref{sim}).}
  \label{eccfig3}
\end{figure}

The case of $a_J = 4.31$~AU is shown in Figure~\ref{eccfig3}, where
the simulation halted after $1.6 \times 10^6$~years. Placing Jupiter
at this particular location has little effect on the orbits of Mars
and Saturn, but Figure~\ref{eccfig3} shows that angular momentum is
transferred from Jupiter to Earth and then subsequently passed to
Venus, as the orbital eccentricities oscillations of Venus and Earth
are directly out of phase with one another. This further demonstrates
that although $a_J = 4.31$~AU does not appear to correspond to a
significant mean motion resonance location (see Figure~\ref{eccfig2}),
the chain of perturbations between Jupiter, Earth, and Venus produces
exceptionally non-linear eccentricity distributions.

\begin{figure}
  \includegraphics[angle=270,width=8.5cm]{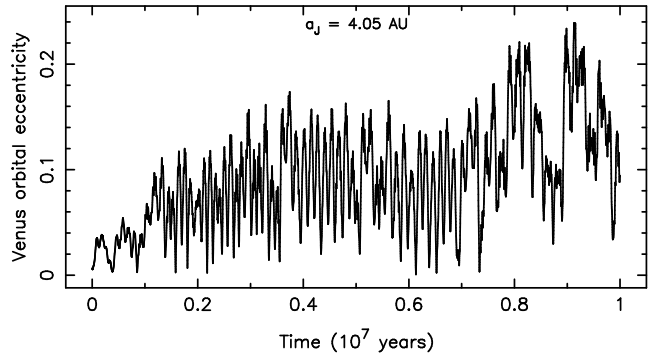}
  \caption{Orbital eccentricity of Venus as a function of time when
    Jupiter is located at a semi-major axis of 4.05~AU.}
  \label{eccfig4}
\end{figure}

As a final example, amongst those simulations for which the solar
system remained stable for the full $10^7$ year duration of our
simulations, the highest values for the Venusian eccentricity occur
when Jupiter is located at $a_J = 4.05$~AU. The results for the
orbital evolution of Venus in this case are shown in
Figure~\ref{eccfig4}. The planetary orbit experiences high-frequency
oscillations and achieves an eccentricity as high as 0.24, but remains
stable nonetheless. Such cases can therefore accommodate a broader
range of Jupiter migration models.

\subsection{Circularization of the Orbit} 
\label{circ}

If Venus once had an orbital eccentricity as high as 0.31, then the
question remains as to how the orbit circularized to its current
state. One of the most efficient mechanisms to circularize a planetary
orbit is through tidal interactions between the planet and its host
star. Using a simple single-planet circularization model based on the
methodology described by \citet{goldreich1966a}, the tidal
circularization timescale is far older than the age of the solar
system. However, numerous other factors, including mutual interactions
between planets, contribute to tidal dissipation that reduces the
tidal circularization timescale \citep{laskar2012}. Hence, to include
both planet-planet interaction and tides in our study, we made use of
the {\sc Posidonius} N-body code
\citep{bolmont2015,blancocuaresma2017b}. {\sc Posidonius} implements
the constant time-lag model (CTL) to account for tidal interactions,
where a planet is modeled as a weakly viscous fluid that is deformed
due to the gradient of the gravitational potential of the central body
(i.e., the host star), \cite[see
  e.g.,][]{mignard1979,hut1981c,eggleton1998a,leconte2010a} The
implementation of this model in {\sc{Posidonius}} has been
successfully used in a number of recent studies \citep[see e.g.,
][]{bolmont2020,nielsen2020b,pozuelos2020}. It is interesting to note
that an alternative tides model exists, namely the constant phase
model (CPL), which may yield faster tidal dissipation timescales
relative to those found via CTL. We refer the reader to
\citet{barnes2017b} and references therein for a detailed review of
existing models used to study the evolution of tides in a planetary
context.

The main free parameters in the CTL model are the degree-2 potential
Love number, $k_2$, and the constant time lag, $\Delta \tau$, of each
considered planet. While the $k_2$ parameter accounts for the
self-gravity and elastic properties of the planet, which can have
values between 0 and 1.5, $\Delta \tau$ represents the lag between the
line connecting the two centers of mass (star--planet), and the
direction of the tidal bulges, which can span orders of magnitudes
\citep{barnes2017b}. In the case of the present-day Earth, the product
of these two parameters (i.e., k$_{2,\oplus}\Delta\tau_{\oplus}$) has
been estimated to be $\sim$213~s \citep{nerondesurgy1997}.

It has been demonstrated that tidal dissipation on present-day Earth
is mostly dominated by friction induced by its seafloor topography on
tidal gravito-inertial waves that propagate in the oceans \citep[see
  e.g.,][]{egbert2000b,egbert2003}. However, on icy-moons such as
Enceladus, tidal dissipation occurs in the sub-surface ocean(s)
\citep[e.g.,][and references therein]{hay2017}. Thus the tidal braking
of an Earth-like planet is mostly dependent on the unknown properties
of any putative oceans, whose properties may evolve over time as a
consequence of losing/gaining oceanic mass, changes in the continental
distribution, seafloor topography, and so forth
\citep{nerondesurgy1997,barnes2017b}. Hence, when exploring the tidal
evolution of a terrestrial exoplanet, the most commonly used approach
is to adopt Earth's current value as a reference for a planet with
surface oceans, where values in the range of (0.1--10)$\times
k_{2,\oplus}\Delta\tau_{\oplus}$ are used to account for different
planetary conditions, such as rocky planets without oceans and those
that are volatile-rich, respectively. This range allows us to explore
a plausible spectrum of values and behaviors \citep[see
  e.g.,][]{bolmont2014,bolmont2020}.

In the solar system, $k_2$ has been measured for several objects,
while $\Delta \tau$ is, in general, poorly known. In the case of
Venus, from the {\it Magellan} mission, the value of $k_2$ was found
to be in the range of 0.23--0.36 \citep{konopliv1996}. Moreover,
\citet{dumoulin2017} computed tidal viscoelastic deformation using
different models of internal structure, and established $k_2$ in the
range of 0.265--0.270 when a solid pure-iron core was considered and
0.27--0.29 when the core was partially or entirely liquid. These
values of $k_2$ are similar to that found for Earth, which was
estimated to be $\sim$0.298 \citep{jagoda2018}. Hence, as a plausible
value for present-day Venus (i.e., a rocky terrestrial planet without
oceans) we assumed $0.1 \times k_{2,\oplus}\Delta\tau_{\oplus}$.

\begin{figure*}
  \begin{center}
    \includegraphics[width=15.0cm]{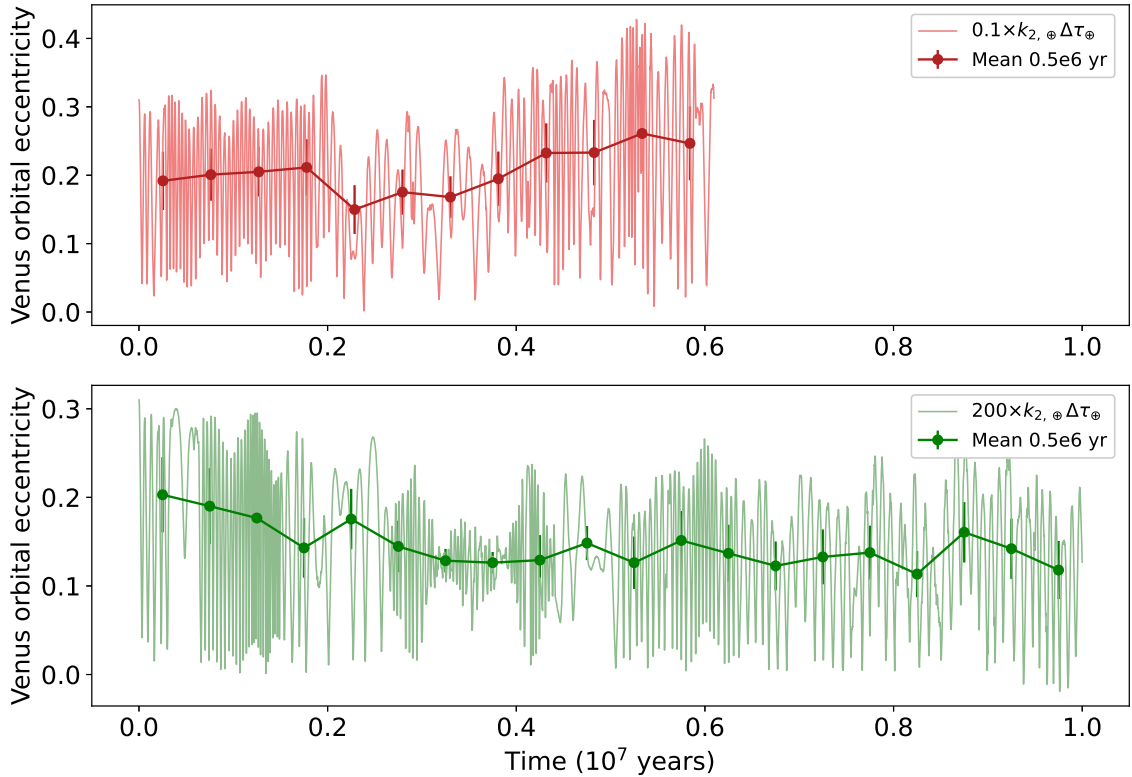}
  \end{center}
  \caption{Venus orbit circularization due to the effects of tides for
    different scenarios: (1) top panel, which considers the current
    Venus tidal dissipation of $0.1 \times
    k_{2,\oplus}\Delta\tau_{\oplus}$; and (2) bottom panel, which
    considers a value of $200 \times
    k_{2,\oplus}\Delta\tau_{\oplus}$. While the current tidal
    dissipation of Venus is insufficient to circularize and stabilize
    the orbit, a large and unrealistic tidal dissipation circularizes
    its orbit in a timescale of $\sim$30~Myr. This latter result
    allowed us to estimate realistic bonds on Venus's past tidal
    dissipation and circularization times of 1.5--5$\times
    k_{2,\oplus}\Delta\tau_{\oplus}$, and 4000--1200~Myr,
    respectively.}
\label{circfig}
\end{figure*}

With all this information, we explored the circularization of the
Venus orbit by performing a suite of simulations considering three
different scenarios within the solar system: (1) initial Venusian
orbital eccentricity excited to 0.31 but with no tidal forces; (2)
initial Venusian orbital eccentricity excited to 0.31 and $0.1 \times
k_{2,\oplus}\Delta\tau_{\oplus}$ (i.e., the plausible current value
representing Venus's tidal dissipation); and (3) initial Venusian
orbital eccentricity excited to 0.31 and $200 \times
k_{2,\oplus}\Delta\tau_{\oplus}$. The scenario (3) tidal dissipation
of $200 \times k_{2,\oplus}\Delta\tau_{\oplus}$ is unrealistic but
accelerates the circularization process and allows us to explore the
possible timescale of circularization due to the effects of
tides. Indeed, as the timescales of the tidal effects scale linearly
with $\Delta\tau$, we are able to estimate the circularization time
for a range of realistic dissipation factors (1-10$\times
k_{2,\oplus}\Delta\tau_{\oplus}$) in a straightforward manner by using
this result \citep{nerondesurgy1997,barnes2017b}. For
self-consistency, we also included the effects of tides for Mercury,
which has a $k_2$ value of 0.45 \citep{noyelles2014}. However, for
simplicity, we did not account for the tidal effects for other planets
beyond Venus. This choice was motivated by the very small tidal
effects experience by terrestrial planets in the Habitable Zone (and
beyond) around solar-like stars \citep{heller2011a}. We integrated the
scenarios described above for $10^7$~yr with a time step of 1~day, and
we accounted for relativistic corrections.

We found that scenario (1) yielded a system which became unstable
within a very short timescale of 2~Myr. This hints that if Venus
reached an eccentricity of 0.31 during its early history, some
mechanism was needed not only to circularize the orbit to reach its
current value but also to stabilize it. In scenario (2), we found that
the system was more stable with respect to scenario (1); however, it
became unstable after 6~Myr. This suggests that the current tidal
dissipation of Venus is not enough to stabilize and circularize its
orbit. In scenario (3), the system was stable for the full integration
time and the orbit started to circularize by decreasing its initial
eccentricity by $\sim$35$\%$. To estimate the circularization time
needed to reach its current value of 0.006, we performed a linear
extrapolation of the eccentricity evolution over $10^7$~yr, finding a
circularization time of $\sim$30~Myr to reach the current value.
Making use of the aforementioned linear-scale property of tides, we
estimated the circularization times for realistic tidal dissipation as
600~Myr for $10 \times k_{2,\oplus}\Delta\tau_{\oplus}$; 1200~Myr for
$5 \times k_{2,\oplus}\Delta\tau_{\oplus}$; 3000~Myr for $2 \times
k_{2,\oplus}\Delta\tau_{\oplus}$; and 6000~Myr for $1 \times
k_{2,\oplus}\Delta\tau_{\oplus}$. These results hint that the current
Earth dissipation is not enough to circularize the Venus orbit on a
timescale that is less than the age of the solar system. However,
these results also clearly show that slightly larger dissipation
factors may be valid. The results corresponding to scenarios (2) and
(3) are shown in Figure~\ref{circfig}.

Taken together, our results suggest that if Venus's eccentricity was
excited up to 0.31 due to Jupiter's migration in the early solar
system, the effects of tides may have played a key role in
circularizing and stabilizing Venus's orbit. We found that the current
value of Venus's tidal dissipation is not enough to achieve this,
suggesting that Venus was not as dry in the past as it is today. To
circularize its orbit over the timescale of the age of the solar
system ($\sim$4000~Myr, post gas phase and migrations), the
dissipation factor needed is $\sim$$1.5 \times
k_{2,\oplus}\Delta\tau_{\oplus}$. This suggests that Venus might have
had a water-rich past, possibly in the form of surface or sub-surface
oceans. Due to the similarities between Venus and Earth, it seems more
realistic to speak about surface oceans, such as those found on Earth,
but we cannot favor any specific location for the water bodies on
ancient Venus due to the limitations of the models used in this
study. This result agrees with the current formation theories of the
solar system, where it has been surmised that Venus and Earth likely
received similar water inventories \citep[see e.g.,][]{raymond2006c}.
Moreover, it has recently been suggested by \citet{way2020} that Venus
may have possessed early oceans that could have created significant
tidal dissipation over $\sim$3000~Myr. Whilst it is not currently
possible definitive to ascertain the true tidal dissipation scenario
experienced by the early Venus, a plausible range of values seems to
be 1.5--5$\times k_{2,\oplus}\Delta\tau_{\oplus}$, which implies
circularization times of 4000--1200~Myr.

There are numerous other processes that also contribute to the
circularization of the Venusian orbit. Tidal dissipation and
circularization are tied to the thermal evolution of the planet,
decreasing the circularization timescale as the planet's interior
cools \citep{driscoll2015}. Furthermore, the interaction between the
atmosphere and surface, strengthened by recent observations of
standing gravity waves, is calculated to be a source of tidal
dissipation for the Venus system
\citep{dobrovolskis1980a,fukuhara2017a}. It should additionally be
noted that since the orbital eccentricity oscillations of Venus
presented here depend on Jupiter's location, subsequent Jupiter
migration can further impact the Venusian orbital variations. For
example, if Jupiter's migration occurred on a timescale longer than
the that of the periodic eccentricity oscillations of the Venus orbit,
then this would act to dampen the amplitude and frequency of the
oscillations. The summation of these sources provides ample
opportunity for the orbit of Venus to arrive at its present state from
the significantly non-circular cases described previously.


\section{Eccentricity Consequences}
\label{con}

The energy budget of the Venusian atmosphere has played a critical
role in the evolution of both atmosphere and surface
\citep{titov2007}. Here, we quantify the impact of the eccentricity
scenarios described in Section~\ref{orbit} on the energy budget of
Venus.


\subsection{Tidal Energy}
\label{tides}

As described in Section~\ref{circ}, tidal dissipation plays a role in
circularizing eccentric orbits with time. The internal heat budget of
the Earth is well known to be sourced primarily from radiogenic
heating and primordial heat. Radiogenic heating of the Earth's
interior, and the resulting heat flux at the surface, can be
calculated using the known mass concentrations of potassium, thorium,
and uranium and their corresponding half-lives
\citep{turcotte2002}. The current total heat flux at the Earth's
surface is $\sim$47~TW \citep{davies2010}, but at 4~Ga the radiogenic
heating alone produced a heat flux of $\sim$80~TW \citep{arevalo2009}.

A third component of internal heat production is gravitational
tides. Tidal heating of planetary interiors and tidal dissipation as a
result of orbital perturbations can result in planets remaining
tidally active for extended periods \citep{renaud2018}. Such heat
dissipation contributes to the surface heat flow budget and thus the
overall energy budget of the atmosphere. Tidal energy potentially
extended the duration of the molten surface scenario for a young
Venus, increasing the time-scale of hydrodynamic escape described by
\citet{hamano2013}. It has been further suggested by
\citet{barnes2009b} that tidal heating of short-period planets can
result in ``Super-Io'' outcomes with significant resurfacing, as is
potentially the case for CoRoT-7b \citep{barnes2010}. Under extreme
circumstances, tidal dissipation can dramatically alter the course of
the atmospheric evolution, such as steering that evolution into a
runaway greenhouse state \citep{barnes2013a}.

To perform our tidal heating calculations, we adopt the methodology of
\citet{jackson2008b,jackson2008c}. Specifically, we use the equation
for the tidal heating rate, $H$:
\begin{equation}
  H = \frac{63}{4} \frac{(G M_\star)^{3/2} M_\star R_p^5}{(3 Q_p / 2
    k_2)} a^{-15/2} e^2
  \label{heatingrate}
\end{equation}
where $G$ is the gravitational constant, $M_\star$ is the stellar
mass, $R_p$ is the planet radius, $Q_p$ is the tidal dissipation
parameter, $k_2$ is the Love number, $a$ is the semi-major axis, and
$e$ is the eccentricity. Note that Equation~\ref{heatingrate} applies
to radial tides, rather than those tides resulting from
non-synchronous rotation. Based on the analysis of {\it Magellan} and
{\it Pioneer Venus} data by \citet{konopliv1996} and the reanalysis of
those data by \citet{dumoulin2017}, we use values of $Q_p = 12$ and
$k_2 = 0.299$. Equation~\ref{heatingrate} shows the dramatic
dependency of the tidal heating on $a$. Thus the distance of Venus
from the Sun results in a relatively low rate of heating due to radial
tides (see Equation~\ref{heatingrate}). Specifically, for the maximum
eccentricity of 0.31 ($a_J = 4.31$~AU scenario) described in
Section~\ref{orbit}, we calculate a tidal heating flux at the surface
of Venus of $\sim$11~GW. The results of the tidal heating calculations
were validated using the tide modules of the Virtual Planet Simulator
\citep{barnes2020}. Although the contribution of the tidal heating to
the total atmospheric energy budget is low, it occurred at a time when
the contribution from the insolation flux was also significantly lower
than that at the present epoch.


\subsection{Insolation Flux}
\label{flux}

The incident (insolation) flux for the solar system planets has
evolved with time as the Sun evolves on the main sequence. Such
evolution can have a profound effect on the time-dependent insolation
flux for terrestrial planets \citep{kulikov2007}, as is seen in
calculations of water-loss for planets around M dwarfs
\citep{luger2015b}. In particular, it is known that the Sun was
$\sim$30\% less luminous in the early era of the solar system
\citep{gough1981d}, during which the radiation environment of Venus
was substantially different.

\begin{figure*}
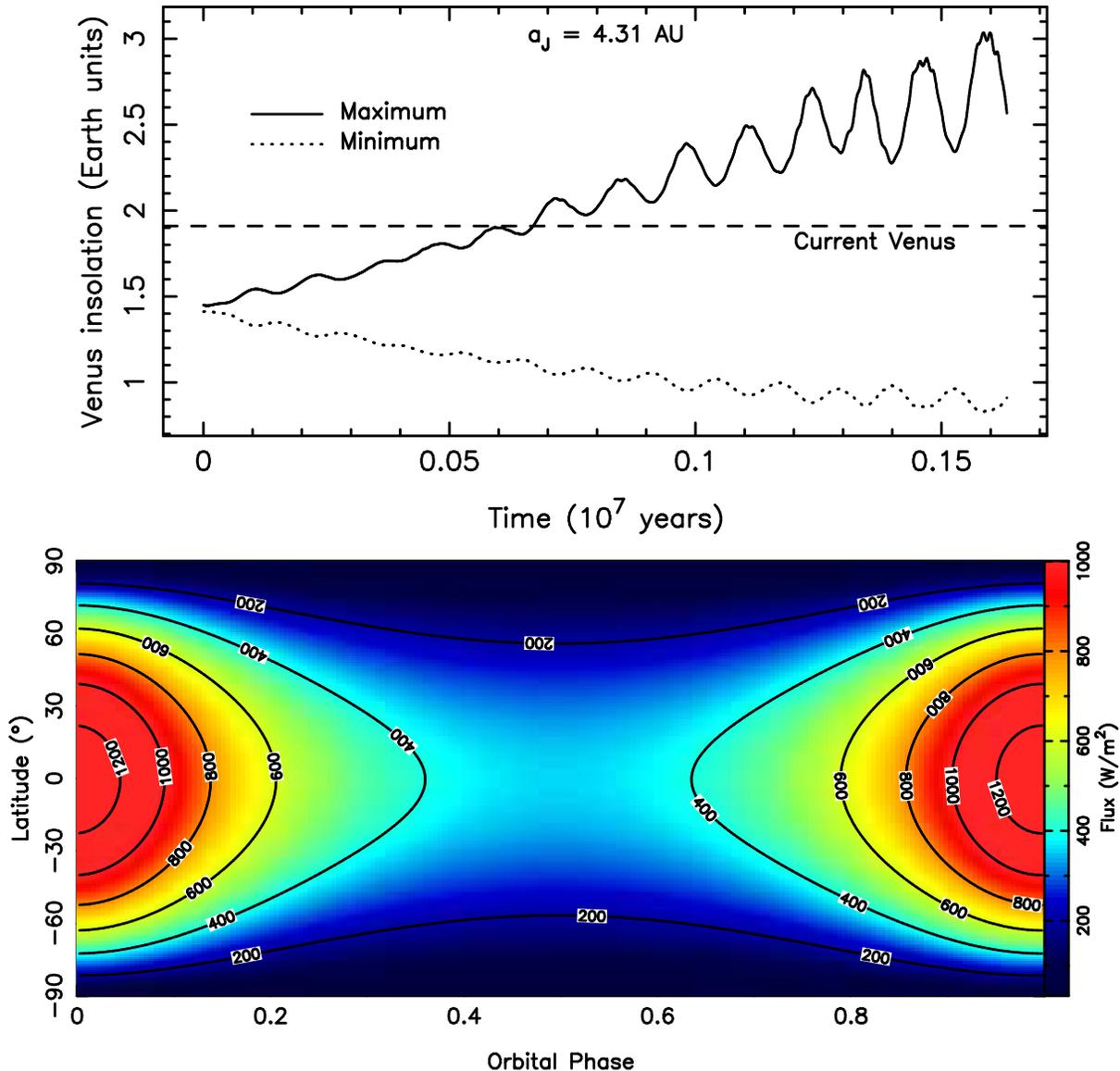

  \begin{center}
    \includegraphics[angle=270,width=14.0cm]{f06a.ps} \\
    \includegraphics[angle=270,width=16.0cm]{f06b.ps}
  \end{center}
  \caption{Top panel: Maximum insolation (perihelion passage) and
    minimum insolation (aphelion passage) at Venus as a function of
    time for the $a_J = 4.31$~AU case (see Figure~\ref{eccfig3}). The
    maximum and mimimum fluxes are represented in units of current
    Earth insolation flux. The horizontal dashed line indicates the
    insolation flux for current Venus. Bottom panel: Diurnal mean
    incident flux intensity map of Venus for an eccentricity of 0.31
    as a function of latitude and orbital phase, where phase zero
    corresponds to perihelion.}
  \label{fluxfig}
\end{figure*}

To simulate the expected insolation flux of Venus during a possible
early era with the high eccentricity described in Section~\ref{orbit},
we adopt a solar luminosity that is 75\% of the current value. At the
semi-major axis of Venus, this results in an insolation flux of $S/S_0
= 1.43$, where $S_0$ is the present-day solar flux received at
Earth. One possible scenario is thus the $a_J = 4.31$~AU scenario (see
Section~\ref{orbit}), where the evolution of the maximum flux
(perihelion) and minimum flux (aphelion) received by Venus is
represented in the top panel of Figure~\ref{fluxfig}. As for
Figure~\ref{eccfig3}, Venus starts in a circular orbit, then the rise
in eccentricity results in a maximum insolation flux that rapidly
starts to oscillate high above its present value, indicated by the
horizontal dashed line.

To demonstrate the effect of the eccentricity on the flux received at
the top of the atmosphere, we simulated the latitudinal flux map of
the planet for the case where the eccentricity reaches 0.31 using the
methodology of \citet{kane2017d}. The flux map is shown in the bottom
panel of Figure~\ref{fluxfig} as a function of orbital phase, where
zero phase corresponds to perihelion passage. We assumed a zero
obliquity for the rotational axis and also that the early Venus was a
rapid rotator prior to its present near spin-orbit
synchronization. The consequence of the assumed rotation means that
the incident flux is calculated as an average at each latitude,
accounting for the reduced flux as a particular longitude rotates away
from maximum solar elevation as well as the zero flux received at the
night side of the planet. As such, the average flux values in the
contour map are significantly lower than the flux received at the top
of the Venusian atmosphere at the sub-solar point. The maximum average
flux received (1297~W\,m$^{-2}$) therefore occurs at the equator
during the perihelion passage. For comparison, the maximum average
flux received by the Earth is 563~W\,m$^{-2}$. Similarly, the minimum
average flux of 360~W\,m$^{-2}$ occurs at the Venusian equator during
aphelion, which is smaller than the minimum average flux received at
the equator of the present Earth during its aphelion
(386~W\,m$^{-2}$).


\subsection{Water Loss}
\label{waterloss}

As described earlier, the loss of water can occur rapidly for planets
orbiting M dwarfs \citep{luger2015b}, and eccentricity-induced tidal
heating can force a runaway greenhouse scenario \citep{barnes2013a}.
Significant water loss can also occur during the early period of
coronal mass ejection for young solar-type stars \citep{dong2017b} and
the incident XUV flux plays an important role in water loss from
CO$_2$ dominated atmospheres \citep{wordsworth2013b}. Here, we combine
several of these aspects to explore the impact of early Venusian
eccentricity on the rate at which the planet would have lost its
water.

\begin{figure}
  \includegraphics[angle=270,width=8.5cm]{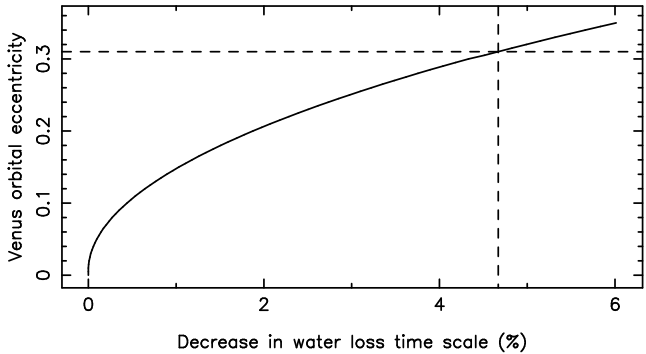}
  \caption{Plot of the decrease in the water loss timescale for the
    full range of Venus eccentricities explored in this study. For the
    maximum eccentricity of 0.31, the timescale to lose one Earth
    ocean of water decreases by 4.7\%, indicated by the dashed lines.}
  \label{waterlossfig}
\end{figure}

A recent study of the water loss implications for eccentric orbits
conducted by \citet{palubski2020a} used 1D climate models to explore
water loss for planets orbiting stars of a range of spectral types. We
utilize the data generated for the G2 main sequence star case, whilst
extending the water loss calculations to high insolation flux
regimes. Shown in Figure~\ref{waterlossfig} are the data for the faint
young Sun scenario described in Section~\ref{flux}, where the
insolation flux at the semi-major axis of Venus is $S/S_0 = 1.43$. The
horizontal axis is expressed in terms of the percentage effect of
eccentricity on the timescale for losing one Earth ocean worth of
water \citep{palubski2020a}. For zero eccentricity, this timescale is
$\sim$3~Gyrs. The dashed lines in Figure~\ref{waterlossfig} highlight
the water loss effect for the eccentricity of 0.31 described in
Section~\ref{orbit}. In this case, the time taken to remove all of the
water is reduced by 4.7\% due to the high eccentricity.

An aspect that is not completely accounted for in the provided water
loss model is the changing luminosity of the Sun. The evolution of the
solar luminosity affects the solar wind received by the planets
\citep{pognan2018}, which in turn can have significant effects on
atmospheric retention \citep{ribas2005,howe2020}. Furthermore, the
pre-main-sequence period of the solar evolution can be characterized
by a relatively high XUV flux that also results in substantial
atmospheric and water loss
\citep{lammer2008,ramirez2014c,gronoff2020b}. As a result, the 4.7\%
reduction in water loss timescale should be considered a lower limit,
and the true water loss rate would probably be substantially higher as
a result of the evolution of the young Sun.

Note that the migration of Jupiter and Saturn likely occurred before
4~Ga, prior to when water would have had an opportunity to be both
substantially delivered and condensed on the surface of Venus and
Earth. However, the tidal circularization timescale estimated in
Section~\ref{circ} demonstrates that the perturbing effect of Jupiter
may have had long-lasting consequences. Thus an increased eccentricity
for Venus could have been sustained long after Jupiter and Saturn
settled to their final (present) locations.


\section{Climate Impacts for Earth and Venus}
\label{climate}

In order to appreciate the possible impacts of eccentricity on Venus,
it is worth briefly reviewing the consequences of eccentricity on
Earth's climate. Earth's eccentricity oscillates roughly between 0.00
and 0.06, with periodicities of about 100 and 400~kyr resulting from
gravitational interactions between Venus (g2), Jupiter (g5), and Mars
(g4) \citep{laskar2004c}. Both numerical calculations and periodic
sedimentary successions found in Earth's geological archive provide
evidence that the 405~kyr eccentricity cycle has been stable for at
least the past 250~Myr \citep{laskar2011a,kent2018}. Despite the small
range of variation, some major climatic, environmental, and biological
perturbations, driven or paced by these eccentricity variations and
the seasonal forcing associated with them, have occurred throughout
Earth's history. For instance, during the glacial-interglacial cycles
of the past million years, the extent of the North Polar ice cap
systematically fluctuated between $80\degr$N and $45\degr$N, with
cycles of 100~kyr \citep{hays1976,ehlers2007}. The waxing and waning
of ice sheets are related to changes in the seasonal insolation at high
latitudes. Geological evidence also demonstrates that eccentricity
modulation on much longer timescales ($\sim$2.4~Myr and $\sim$9~Myr)
has led to the systematic release and sequestration cycles of organic
carbon \citep{boulila2012,martinez2015,batenburg2016,laurin2016} which
in turn impacted atmospheric greenhouse gas concentrations and, in
some cases, possibly led to major perturbation events during which
ocean bottom water fell anoxic, leading to the extinction of several
marine species \citep{kuhnt2005,devleeschouwer2017c}. More often than
not, astronomically forced perturbations to the Earth's system were
related to biological processes impacting the carbon cycle, but the
strength of seasonality (i.e., the combined effect of precession and
obliquity enforced by eccentricity effects) also plays a crucial role
in many other Earth system processes, such as atmospheric heat
transport \citep{donohoe2013a} and ocean dynamics
\citep{lisiecki2014}.

The potential variations of the early Venusian eccentricity may have
had severe consequences for the climate evolution. Although the tidal
energy effects are relatively low (Section~\ref{tides}), the
insolation flux effects (Section~\ref{flux}) and resulting water loss
(Section~\ref{waterloss}) may have transferred sufficient water to the
atmosphere to accelerate the development of a moist greenhouse
\citep[e.g.,][]{kasting1984a,gomezleal2018}. Although CO$_2$ alone can
lead to a non-reversible runaway greenhouse state \citep{popp2016},
the water distribution between the surface and the atmosphere plays an
important role. The insolation flux for the early Venus considered
here is 1.43 times the current solar constant, and
\citet{kasting1988c} predicted that the insolation flux at which
complete water loss occurs is 1.4 times the current solar
constant. However, \citet{wolf2014a}, using 3D models rather than the
1D models of \citet{kasting1988c}, argued that the onset of a moist
greenhouse can occur at significantly higher insolation fluxes. Even
so, the water loss rates presented in Section~\ref{waterloss} indicate
a $\sim$5\% reduction in the time to transfer an Earth ocean worth of
water to the atmosphere due to enhanced eccentricity, resulting in
substantial atmospheric water in the early Venusian atmosphere. For
example, $\sim$$10^{-3}$\% of Earth's current surface water inventory
resides in the atmosphere. The effect of increasing the atmospheric
water content, even by a small amount, can have stochastic climate
consequences, depending on the insolation flux and atmospheric
composition and structure \citep{goldblatt2015}. One scenario for a
high Venusian atmospheric water content is that it leads to an
acceleration of a moist greenhouse, followed by photodissociation
\citep{zahnle1986b,kulikov2006,kasting2015,lichtenegger2016}. Alternatively,
it has been argued that the moist greenhouse phase does not always
occur, transitioning directly into a runaway greenhouse
\citep{leconte2013c,ramirez2020b}.


\section{Implications for Exoplanets}
\label{implications}

The architectures of exoplanetary systems is a topic of rapidly
advancing research, particularly in light of the expanding inventory
and demographics of exoplanet discoveries
\citep{ford2014,winn2015}. Smaller (terrestrial) planets tend to have
smaller eccentricities, particularly those in compact systems
\citep{kane2012d}. However, the work described here shows that the
early stages of architecture evolution may be more turbulent in the
presence of a migrating giant planet. Thus giant planets in
exoplanetary systems may play a critical role in the development of
terrestrial planet climates.

Studies of giant planet occurrence rates at long orbital periods have
demonstrated that their frequency is relatively low
\citep{hill2018,wittenmyer2020b}. Furthermore, observational evidence
points toward giant planets being even scarcer around M dwarf stars
\citep{cumming2008,johnson2010d}. If giant planet migration processes,
similar to that undertaken by Jupiter and Saturn, are a common feature
of early planetary system development, then the excitation and
subsequent dampening of inner terrestrial planet eccentricities may be
common in such systems. The profound implication is that the presence
of giant planets in the outer part of planetary systems may increase
the likelihood of water loss and runaway greenhouse climate evolution,
resulting in a more extensive Venus Zone than previously estimated
\citep{kane2014e}. As the sensitivity of exoplanet measurements
continue to increase in the long-period regime, and as the atmospheres
of Venus analogs are detected, this proposed correlation may be
properly evaluated \citep{lincowski2019,lustigyaeger2019b}.


\section{Conclusions}
\label{conclusions}

The evolution of the Venusian climate still has many outstanding
questions, most particularly as to whether the planet was habitable
until relatively recently \citep{way2016} or was devoid of surface
liquid water from its formation \citep{hamano2013}. Answering these
questions is crucial for understanding not just the comparative
evolution of Earth and its sibling planet, but the prevalence and
sustainability of planetary habitability in a broad range of
exoplanetary systems. There have been many differences between Venus
and Earth with regard to their overall planetary evolution, and
determining the dominant factors that drove those evolutions remains a
topic of active research.

The study presented here specifically investigates the effect of
possible orbital dynamical scenarios on the evolution of an early
Venus. Our simulations and subsequent analyses demonstrate that (1)
the eccentricity of the Venusian orbit is dramatically increased for
particular locations of Jupiter and (2) the consequences of the
increased eccentricity would have included a significantly increased
rate of surface liquid water loss. Clearly there exists a vast
parameter space of initial orbital conditions that are possible, of
which the study presented here has explored a subset. Furthermore, our
simulations do not include the migration effects of Jupiter and Saturn
during the course of individual dynamical simulations. The inclusion
of specific migration patterns would marginalize the results given the
plethora of possible migration scenarios (see
Section~\ref{migration}). Rather, our methodology considers each
simulation as an encapsulated exploration of the perturbative effects
of Jupiter as a function of semi-major axis to provide a first-order
picture of possible dynamical scenarios during migrational
pathways. However, our investigations of tidal dissipation and
circularization timescales show that damping the eccentricity
perturbations of Venus to their current value requires a larger
initial water inventory than that for the current Earth, lending
credence to the notion of substantial water delivery to an early
Venus.

If Jupiter did indeed accelerate the early climate development of
Venus, then similar situations may have occurred elsewhere. The
occurrence rate of detectable giant planets near or beyond the snow
line is known to be relatively low. However, for those systems with
outer giant planets, the early dynamical history of those systems may
cause a measurable increase in the occurrence rate of inner
terrestrial planets in runaway greenhouse states. Such dynamical
impacts on terrestrial climate evolution are thus an important factor
in considering planetary evolution and habitability.


\section*{Acknowledgements}

The authors would like to thank Sean Raymond for detailed feedback
regarding migration and formation models, Cayman Unterborn for
discussions regarding planetary interiors, Igor Palubski for
communications regarding water loss models, and Anthony Del Genio for
feedback regarding climate evolution. The authors would also like to
thank the anonymous referees for their insightful comments on the
manuscript. P.V. acknowledges funding support from the Heising-Simons
Foundation. The results reported herein benefited from collaborations
and/or information exchange within NASA's Nexus for Exoplanet System
Science (NExSS) research coordination network sponsored by NASA's
Science Mission Directorate.


\software{Mercury \citep{chambers1999}, VPLanet \citep{barnes2020},
  Posidonius N-body \citep{blancocuaresma2017b}}




\end{document}